\documentclass[twocolumn,amsmath,amssymb]{article}

\makeatletter
\def\@normalize{@setsize\normalsize{12pt}\xpt\@xpt
\abovedisplayskip 10pt pluse2pt minus5pt\belowdisplayskip
\abovedisplayskip \abovedisplayshortskip \z@ plus3pt\belowdisplayshortskip
6pt plus3pt minus3pt\let\@listi\@listI}
\def\subsize{\@setsize\subsize{12pt}\xipt\@xipt}
\def\section{\@startsection{section}{1}{\z@}{24pt plus 2 pt minus 2 pt} {12pt plus 2pt
minus 2pt}{\large\bf}}
\def\subsection{\@startsection{subsection}{2}{\z@}{12pt plus 2 pt minus 2 pt}{12pt plus
2pt minus 2pt}{\subsize\bf}}
\makeatother

\usepackage{graphicx} 
\usepackage{dcolumn} 
\usepackage{bm} 

\begin{document}
\date{Nov 15, 2006}
\title{\large\bf Local cause of coherence in Boolean networks}
\author{Chikoo Oosawa$^{1,}\footnote{Correspoding author: chikoo@bio.kyutech.ac.jp}$ $^{,}$\footnote{ Also at: Bioalgorithm Project, Faculty of Computer Science and Systems Engineering, Kyushu Institute of Technology} , Kazuhiro Takemoto$^{2}$, Shogo Matsumoto$^{3}$, Michael A. Savageau$^{4}$\\\\$^{1}$Department of Bioscience and Bioinformatics, Kyushu Institute of Technology \\ Iizuka, Fukuoka 820-8502, Japan \\ $^{2}$Bioinformatics Center, Institute for Chemical Research, Kyoto University, \\ Gokasho, Uji, Kyoto 611-0011, Japan \\$^{3}$Graduate School of Computer Science and Systems Engineering, \\ Kyushu Institute of Technology, Iizuka, Fukuoka 820-8502, Japan \\$^{4}$Department of Biomedical Engineering, College of Engineering, \\University of California, Davis 95616, U.S.A.}

\maketitle
\thispagestyle{empty}
\subsection*{\centering Abstract}
\vspace*{-3mm}
We performed a numerical study on random Boolean networks with power-law rank outdegree distributions to find local structural cause for the emergence of high or low degree of coherence in binary state variables. The degree of randomness and coherence of the binary sequence are measured by entropy and mutual information, depending on local structure that consists of a node with a highly connected, called hub and its upstream nodes, and types of Boolean functions for the nodes. With a large number of output connections from a hub, the effects of Boolean function on the hub are more prominent. The local structures that give larger entropy tend to give rise to larger mutual information. Based on the numerical results and structural conditions we derived a time-independent transmission characteristic function of state variables for  the local structures. We obtained good relationships
between the numerical and analytical results, which indicate that dynamical properties from the whole networks can be inferred from the differences in the local structures.\\
\\keywords: Boolean networks; power-law; coherence; mutual information; entropy; transcriptional regulatory networks

\section{Introduction}
Biological system consists of complex adaptive systems, for example, neural and transcriptional regulatory systems whose structures often can be abstracted to a graph or network. In general, the systems perform their functions
correctly when certain appropriate communications among nodes are established, because such systems need to add or delete own nodes, or change the strength of connectivity to optimally adapt to exogenous inputs. The transcriptional regulatory network is one of the complex adaptive systems where the node mainly corresponds to the transcriptional unit, and responds to the environmental changes to survive and proliferate. The Boolean network\cite{ORIGIN} is one of the discrete dynamical models for the transcriptional regulatory network and exhibits binary sequences of the state variables that represent expression pattern of the transcriptional regulatory network. Since the state variables in the Boolean network are sensitive to inputs from other nodes via directed edges, and affect other nodes , quality of communication is characterized as the size of mutual information. The mutual information indicates the degree of coherence, synchronization, amount of information content in the state variables, or potential for computational capability of the network\cite{LANGTON}.
\\
\quad In this study, we show local structural cause for the emergence of coherence in a Boolean network. Since we embedded power-law rank output connectivity distributions in the Boolean networks whose input connectivity $K_{in}$= 2, the model networks have some hubs that integrate many output connections. Because the hubs synchronously transmit their state to the downstream nodes, the downstream nodes are affected by single or multiple hubs simultaneously. The structural condition seems to automatically provide global coherence in the state variables; however, the structural aspects give only possible influence from the hubs. In fact, we need to consider a type of Boolean function at  the hubs and their upstream nodes. We show both the effects of Boolean functions and the number of output connections on the size of entropy and mutual information. In addition,  some results can be explained by transmission characteristic functions that are derived from the numerical results and local structures.\\\\
\quad Please note that Tables \ref{tab:ar1}--\ref{tab:cmb2} and Figs. \ref{fig:3nodes}--\ref{fig:tcf} can be found in Sec. \ref{sec:app}, Appendix.

\section{Model}
Dynamics of the Boolean networks\cite{ORIGIN,COMAS} are determined by 
\begin{equation}
X_i(t+1)=B_i\left[\bm{X}(t)\right]\quad(i=1,2,...,N),
\label{eq:bn}
\end{equation}
where $\bm{X}(t)$ is the binary state, either 0 or 1, of node {\it i} at time $t$; $B_i(\cdot)$ is the Boolean function [see Table \ref{tab:bool}] used to simultaneously update the state of node {\it i}; and $\bm{X}(t)$ is a binary vector that gives the states of the $N$ nodes in the network. After assigning initial states $\bm{X}(0)$ to the nodes, the successive states of the nodes are updated by input states and its Boolean function. The dynamical behavior of these networks is represented by the time series of the binary states. The time course follows a transient phase from an initial state until a periodic pattern, known as an attractor, is eventually established.

\begin{table}[hbt]
\caption{Four of 16 Boolean functions with indegree $K_{in}$= 2. In this paper, only 1, 2, 4, and 8 are used because of the feasibility and biological meaning of the functions\cite{JTB,PNAS}.}
\begin{center}
\begin{tabular}{|cc|*{4}{r}|}\hline
\multicolumn{2}{|c|}{Input}&\multicolumn{4}{c|}{Output}\\ \hline
0&0&0&0&0&1\\
0&1&0&0&1&0\\
1&0&0&1&0&0\\
1&1&1&0&0&0\\
\hline\hline
Decimal&index&\enspace 1&\enspace 2&\enspace 4&\enspace 8\\
\hline
\end{tabular}\label{tab:bool}
\end{center}
\end{table}

\section{Numerical Condition}
\begin{figure}[h]
\begin{center}
\includegraphics[scale=0.35]{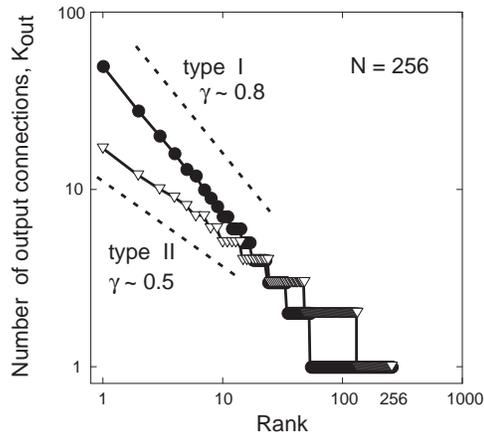}
\caption{Power-law rank connectivity distributions in the model. Power exponent, $\gamma$ for type I and II is about 0.8 and 0.5, respectively, where $K_{out}(rank) \sim rank^{-\gamma}$. We performed only a single network size of 256 in this paper.}
\label{fig:topo}
\end{center}
\end{figure}

We constructed $10^4$ Boolean networks in each power-law rank distribution [see Fig. \ref{fig:topo}] with fixed network size. 2$\times10^3$ initial states were applied to each network. Four different Boolean functions [see Table \ref{tab:bool}] were used in almost equal probability [see Table \ref{tab:ar1}]. Please note that all the generated networks use the same amount of resources since the size of network is fixed--256 nodes and 512 directed edges.

\begin{figure}[h]
\begin{center}
\includegraphics[scale=0.6]{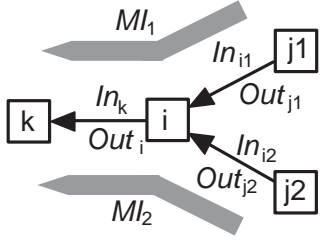}
\caption{Flow of state variables from upstream to downstream. Since input connectivity for all nodes is equal to 2 (=$K_{in}$), there are two  pathways for mutial information in each node. Input sequence $In_{i1}$ for the node $i$ is the same as the output sequence of an upstream node $Out_{j1}$, and the output sequence $Out_i$ for the node $i$ is the same as the input sequence of a downstream node $In_k$. When node $i$ has multiple output connections, it has the same binary sequence because state variables in networks are subject to Eq. (\ref{eq:bn}).}
\label{fig:3nodesMI}
\end{center}
\end{figure}

\quad We measured entropy (randomness) and mutual information (coherence) of the state variables to characterize the dynamics of the Boolean networks\cite{LANGTON,COMAS} [see Fig. \ref{fig:3nodesMI}].

\section{Results}
\subsection{Numerical results}
\label{sec:nr}
\begin{figure}
\begin{center}
\includegraphics[scale=0.254]{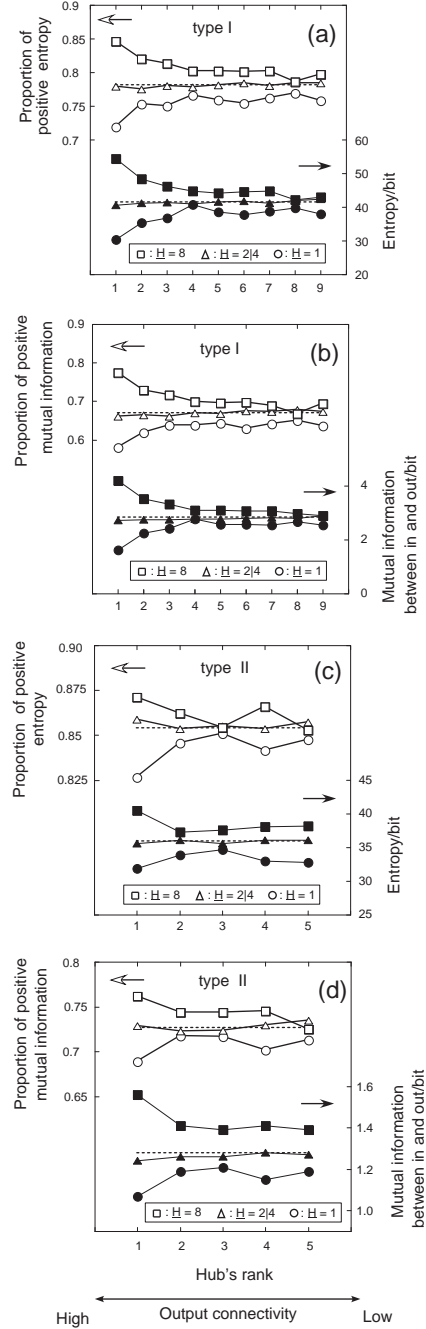}
\caption{Dependence of proportion of positive entropy (a) and (c) and mutual information (b) and (d) on type I and II distribution in Fig. \ref{fig:topo}. Three different symbols denote different types of Boolean function on hubs; square: type 8, triangle: type 2 or 4, and circle: type 1 [see Table \ref{tab:bool}]. The vertical bar "$|$" in the legend denotes either of the Boolean function on each side of the bar. \underline{H} denotes the hub's Boolean function.  Dashed horizontal lines indicate the proportions from all realizations [see Table \ref{tab:ar1}].}
\label{fig:rankT1}
\end{center}
\end{figure}

In total, we obtained 70622 and 177098 attractors from type I and type II distribution, respectively. Size of entropy and mutual information are measured from the attractors. Together with the network structural condition, we show the dependence of distribution's rank on entropy and mutual information in Fig. \ref{fig:rankT1}.\\ 
\quad The dependence of the Boolean function on entropy and mutual information is prominent in higher ranked hubs on both the output distribution styles. The Boolean functions that give larger entropy tend to give larger mutual information. These results suggest that the collective (global) coherence in the state variables of the networks may be due to the style of upstream (local) conditions, including connections among the upstream nodes and assignment of the Boolean functions.

\subsection{Transmission characteristic function}
\label{sec:tcf}
\begin{figure}[h]
\begin{center}
\includegraphics[scale=0.351]{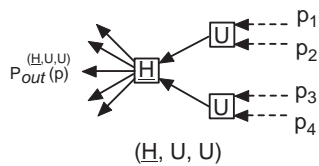}
\caption{Typical local structure around a hub. Squares and directed edges (arrows) correspond to the nodes and connections of binary sequence pathways, respectively. Dashed arrows indicate inputs from upstreams. \underline{H} and U in the squares correspond to the Boolean function on hub and its upstream nodes. Each node has one of the Boolean functions, 1, 2, 4, and 8 as shown in Table \ref{tab:bool}. A set of three characters in a brace at the bottom of each figure denotes a combinatorial set of the Boolean function on nodes in the local structure. $P^{(\underline{H},U,U)}_{out}(p)$ represents transmission characteristic function of input probability $p$ [see Fig. \ref{fig:tcf}] and the combination of the Boolean functions [see Table \ref{tab:cmb2}].}
\label{fig:3nodes}
\end{center}
\end{figure}

\begin{figure}[h]
\begin{center}
\includegraphics[scale=0.30]{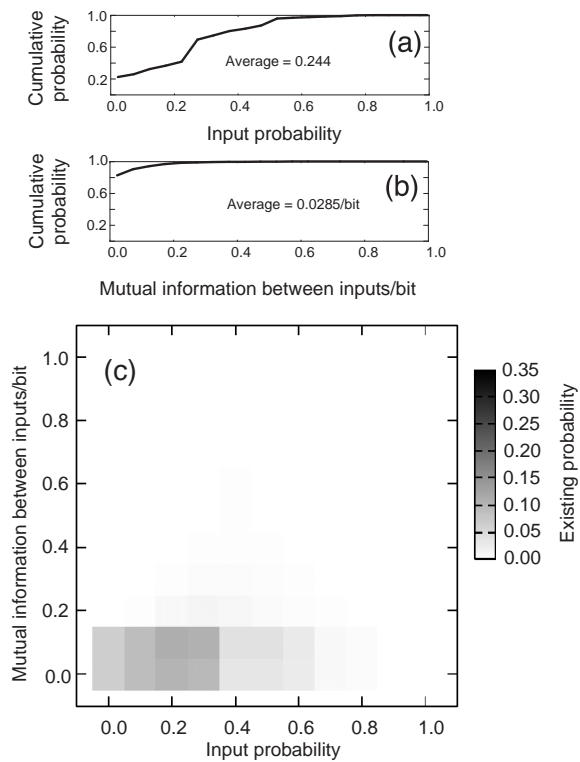}
\caption{Statistical properties of inputs to the local structure in Fig. \ref{fig:3nodes}: (a) Cumulative distribution of input probability($p_1-p_4$). (b) Cumulative mutual information between 4 inputs. (c) Correlation between input probability (a) and mutual information (b). These distributions are obtained from 1st-ranked hubs on the basis of 70622 attractors from $10^4$ networks with style I distribution and 2$\times10^7$ initial states.}
\label{fig:uncorr}
\end{center}
\end{figure}

\begin{figure}[h]
\begin{center}
\includegraphics[scale=0.30]{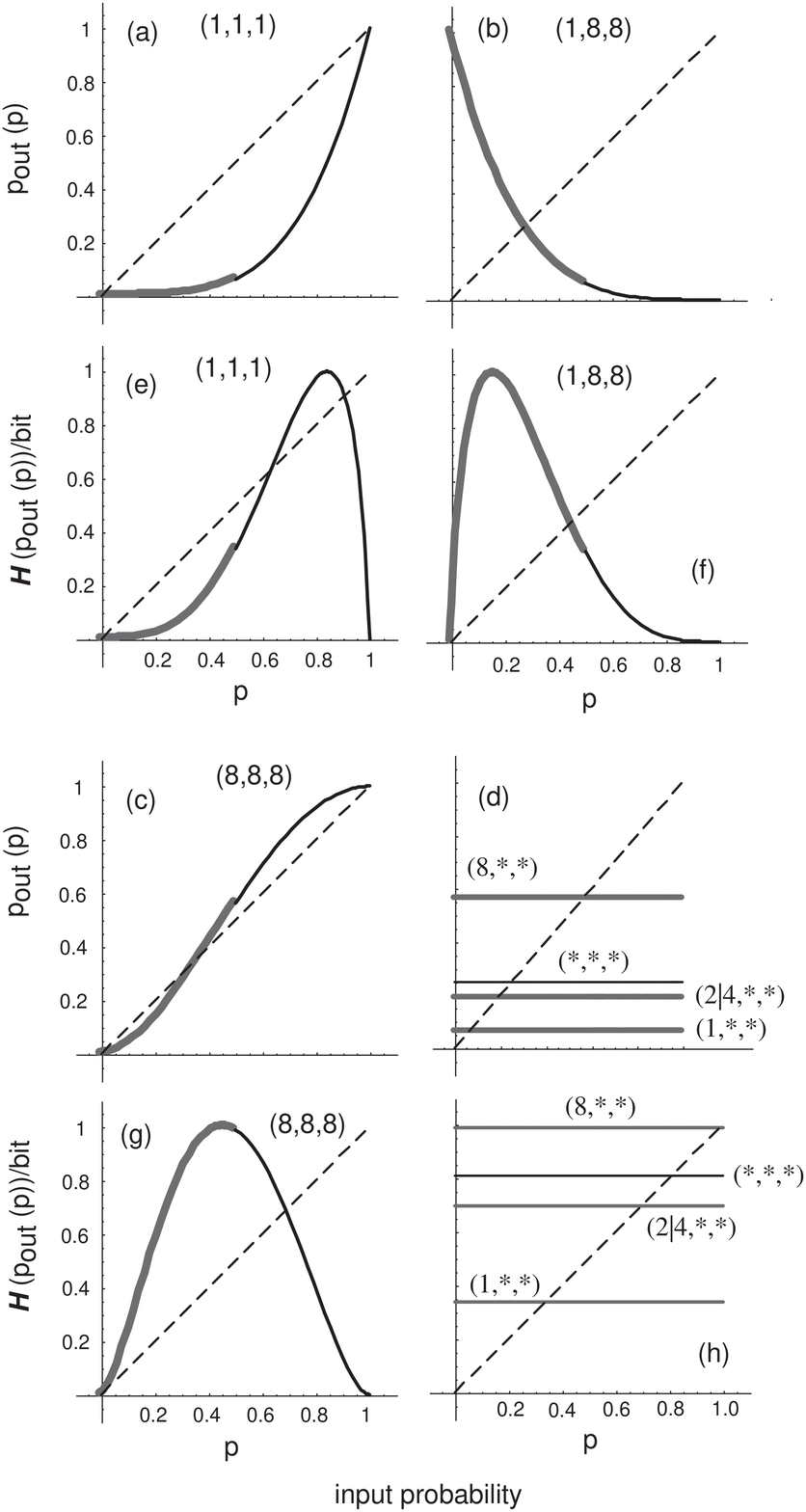}
\caption{Example of transmission characteristic functions (a)--(d) and their output entropy (e)--(h). Each set of 3 characters in a brace shows a Boolean function combination [see Table \ref{tab:bool} and Fig. \ref{fig:3nodes}]. A vertical bar "$|$" denotes either of the Boolean functions on each side of the bar. An asterisk "*" denotes any of the Boolean functions in Table \ref{tab:bool}. The triple asterisks, (*,*,*) indicates all the 18 possible combinations. (a) Transmission characteristic function for (1,1,1), $P_{out}^{(1,1,1)}(p)=p^4$. Other transmission characteristic functions are also described as at the most of 4th order polynomial. (c) Entropy of the transmission characteristic function in (a). (d) Average transmission characteristic functions as indicated in braces. (h) Entropy of the transimission characteristic function in (d). As a guide for (a)--(d), thick region of these curves correspond to $p\in[0, 0.5]$. Dashed lines show $P_{out}(p)=p$ or $H(P_{out}(p))=p$ relationship for (e)--(h).}
\label{fig:tcf}
\end{center}
\end{figure}

To elucidate dependence of rank on randomness and coherence, we first focus on the local structure around the hubs as shown in Fig. \ref{fig:3nodes}. Because each node has two inputs by definition, outputs from the hubs are subjected to 4 inputs and 3 Boolean functions. From numerical results, we obtain statistical properties to determine the relationships between the 4 inputs. Figure \ref{fig:uncorr} shows two statistical properties: (1) Range of input probability is almost limited from 0.0 to 0.5. (2) Degree of correlations between inputs is very low in the range of input probability.\\
\quad The two statistical properties allow us to take the following analytical approach. We define the output property as a function of the input probability, called transmission characteristic function [see Fig. \ref{fig:tcf}]. For simplicity, we assume that the 4 inputs ($p_1-p_4$ in Fig. \ref{fig:3nodes}) receive binary sequence with the same probability\cite{COKT}. The transmission characteristic functions can be described as at the most of 4th order polynomial and obtained by combining Boolean functions [Table \ref{tab:cmb2}] on the local structure [Fig. \ref{fig:3nodes}].\\
\quad When we ignore the specificity of the Boolean functions in upstream nodes in a local structure, we can obtain the average transmission characteristic function-considering only the hub's Boolean function, weighted average of the individual transmission characteristic functions can be obtained. The average transmission characteristic function can be written as

\begin{eqnarray}
\left(P_{out}^{(1,*,*)}(p)=\frac{1}{16}\right)<\left(P_{out}^{(2|4,*,*)}(p)=\frac{3}{16}\right)\nonumber \\ <\left(P_{out}^{(*,*,*)}(p)=\frac{1}{4}\right)<\left(P_{out}^{(8,*,*)}(p)=\frac{9}{16}\right), 
\label{eq:ap}
\end{eqnarray}

where an asterisk "*" denotes any one of the Boolean functions in Table \ref{tab:bool}. The triple asterisks in the braces represent the average transmission characteristic function from the over all 18 [see Table \ref{tab:cmb2} and Fig. \ref{fig:tcf}] Boolean function combinations. The order of their entropies is described as

\begin{eqnarray}
\bm{H}\left(P_{out}^{(1,*,*)}(p)\right)<\bm{H}\left(P_{out}^{(2|4,*,*)}(p)\right) \nonumber \\<\bm{H}\left(P_{out}^{(*,*,*)}(p)\right)<\bm{H}\left(P_{out}^{(8,*,*)}(p)\right).
\label{eq:ae}
\end{eqnarray}

The same order of entropy size is already shown in Fig. \ref{fig:rankT1} (a) and (c). In particular, the tendency can be seen in higher ranked hubs in both styles I and II, demonstrating that the transmission characteristic functions are useful tools to infer network dynamics.

\section{Summary and Discussion}
On comparing the average transmission characteristic functions and numerical results, we obtained good relationships between them, suggesting that the local structure at the hubs affects global dynamical properties in the networks. These results also provide a blueprint of design principle for an artificial gene regulatory network, and help to elucidate the role of hubs in dynamical system.\\
\quad In recent years, many large-scale complex networks such as social, metabolic, and neural networks have been paid great attention from many fields. It has been revealed that the biological networks contain prominent local structures, called motif, subgraph or clique that consist of a small number of nodes and edges \cite{MOTIF,PNAS-AB}. In this paper, we demonstrate just one of the local structures in simplified dynamical model for the transcriptional regulatory networks. In fact, size of the biological complex adaptive systems varies over a very long time because of many various events. Since clique models\cite{KTCO1,KTCO2} seem to be a promising approach for explaining growing complex networks, an exchange of information among local structures may contribute to the mechanisim for maintenance and growth of complex networks.

\subsection*{Acknowledgments}
This work was partially supported by Grant-in-Aid No. 18740237 from MEXT (JAPAN) and the US Department of Defence Grant N00014-97-1-0364 from the Office of Naval Research.

\section{Appendix}
\label{sec:app}
\begin{table}[h]
\caption{Numbers of realizations out of $10^{4}$ networks are indicated. Four different Boolean functions [see Table \ref{tab:bool}] at the hubs are used with equal probability. "\underline{H}" in the table denotes the hubs Boolean function [see Fig.\ref{fig:3nodes}].}
\begin{center}
\begin{tabular}{|c||c|c|c|c|}
\hline
Hub's rank&\underline{H} = 1&\underline{H} = 2&\underline{H} = 4&\underline{H} = 8\\
\hline
1&2496&2508&2540&2456\\
2&2564&2517&2459&2460\\
3&2463&2452&2564&2521\\
4&2575&2471&2518&2436\\
5&2454&2556&2529&2461\\
6&2516&2467&2476&2541\\
7&2525&2609&2486&2380\\
8&2502&2494&2491&2513\\
9&2529&2519&2445&2507\\
\hline
\end{tabular}\label{tab:ar1}
\end{center}
\end{table}

\begin{table}[h]
\caption{The combination number of the Boolean functions on the local structure shown in Fig. \ref{fig:3nodes}. 64 $(=4^3)$ combinations can be reduced to 18 because of the symmetry in the local structure [see Fig. \ref{fig:3nodes}] and in-out relations in the Boolean functions [see Table \ref{tab:bool}]. There are two major columns, and each major column consists two subcolumns. The left subcolumns contain all possible combinations of the Boolean functions in the case that hub has one of the Boolean funtions. The right subcolumns indicate integrated combinations. The combinations with dagger can be consolidated with the combinations with prime. A set of numbers at both ends of the arrow at the bottom line in each major column denotes the original and reduced number of combinations. When Boolean function of 8 is located in the hub, the same procedure as shown in left major column can be taken.}
\begin{center}
\begin{tabular}{|*{1}{@{\tiny}c|}*{1}{@{\tiny}c||}*{2}{@{\tiny}c}|*{1}{@{\tiny}c|}}\hline
\multicolumn{2}{|c||}{(1,U,U)}&\multicolumn{3}{c|}{(2,U,U) or (4,U,U)}\\ \hline
(1,1,1)&(1,1,1)&(2,1,1)$^{\dag}$&(4,1,1)$^{\dag}$&(2$|$4,1,1)$\prime$\\ 
(1,1,2)$^{\dagger}$&(1,1,2$|$4)$\prime$&(2,1,2)&(4,1,2)&(2$|$4,1,2$|$4)\\ 
(1,1,4)$^{\dagger}$&(1,1,8)&(2,1,4)&(4,1,4)&(2$|$4,1,8)\\ 
(1,1,8)&(1,2$|$4,2$|$4)&(2,1,8)&(4,1,8)&(2$|$4,2$|$4,2$|$4)\\ 
(1,2,1)$^{\dagger}$&(1,2$|$4,8)&(2,2,1)&(4,2,1)&(2$|$4,2$|$4,8)\\ 
(1,2,2)&(1,8,8)&(2,2,2)&(4,2,2)&(2$|$4,8,8)\\ 
(1,2,4)&	&(2,2,4)&(4,2,4)&\\ 
(1,2,8)&	&(2,2,8)&(4,2,8)&\\ 
(1,4,1)$^{\dagger}$&	&(2,4,1)&(4,4,1)&\\ 
(1,4,2)&            &(2,4,2)&(4,4,2)&\\ 
(1,4,4)&            &(2,4,4)&(4,4,4)&\\ 
(1,4,8)&            &(2,4,8)&(4,4,8)&\\ 
(1,8,1)&            &(2,8,1)&(4,8,1)&\\ 
(1,8,2)&            &(2,8,2)&(4,8,2)&\\ 
(1,8,4)&            &(2,8,4)&(4,8,4)&\\ 
(1,8,8)&            &(2,8,8)&(4,8,8)&\\ 
\hline\hline
\multicolumn{2}{|c||}{16 $\Longrightarrow$ 6}&\multicolumn{3}{c|}{32 $\Longrightarrow$ 6}\\ \hline
\end{tabular}\label{tab:cmb2}
\end{center}
\end{table}

\end{document}